\documentclass[longauth, traditabstract]{aa} 
\newcommand{\Av}{{A$_V$}}
\newcommand{\cmsq}{{cm$^{-2}$}}

\newcommand{\mstar}{M$_\odot$}

\newcommand{\um}{$\mu$m}

\newcommand{\herschel}{{\it Herschel}}
\newcommand{\spitzer}{{\it Spitzer}}
\newcommand{\Planck}{{\it Planck}}
\newcommand{\partemis}{{P-ArT\'{e}MiS}}
\newcommand{\artemis}{{ArT\'{e}MiS}}
\newcommand{\mpul}{{M$_{\rm line}$}}

\usepackage{natbib}
\bibpunct{(}{)}{;}{a}{}{,} \usepackage{graphicx, pstricks}
\usepackage{mathtools,amsmath}
\usepackage{wasysym,afterpage}
\usepackage{rotating}
\usepackage{txfonts}

\begin{document}
\title{Resolving the Vela~C ridge 
with \partemis\thanks{This publication is based on data acquired with the Atacama 
Pathfinder Experiment (APEX) in ESO program 083.C-0996. 
APEX is a collaboration between the Max-Planck-Institut f\"ur Radioastronomie, the European
Southern Observatory, and the Onsala Space Observatory.} 
and \herschel\thanks{\emph{Herschel} is an ESA space observatory with science
instruments provided by European-led Principal Investigator
consortia and with important participation from NASA.} }
\titlerunning{}
\authorrunning{T. Hill et al.}
        \author{T. Hill\inst{1}
          \and Ph. Andr\'e\inst{1}
          \and D. Arzoumanian\inst{1}
            \and F. Motte\inst{1}
          \and V. Minier\inst{1}          
          \and A. Men'shchikov\inst{1}
                    \and P. Didelon\inst{1}, M. Hennemann\inst{1}, V.~K\"onyves\inst{1,2}, Q. Nguy$\tilde{\hat{\rm e}}$n-Lu{\hskip-0.65mm\small'{}\hskip-0.5mm}o{\hskip-0.65mm\small'{}\hskip-0.5mm}ng\inst{3}, P. Palmeirim\inst{1} 
                    \and  N. Peretto\inst{1}, N. Schneider\inst{1,4,5}, S. Bontemps\inst{4,5}, F. Louvet\inst{1} 
                   \and D. Elia\inst{6}, T.~Giannini\inst{6}
       \and V.~Rev\'eret\inst{1},  J.~Le~Pennec\inst{1}, L.~Rodriguez\inst{1}, O.~Boulade\inst{1}, E.~Doumayrou\inst{1}, 
       D.~Dubreuil\inst{1}, P.~Gallais\inst{1}, M.~Lortholary\inst{1}, J.~Martignac\inst{1}, M.~Talvard\inst{1}, C.~De~Breuck\inst{7}
        }
   \institute{Laboratoire AIM, CEA/IRFU CNRS/INSU Universit\'e Paris Diderot, CEA-Saclay, 91191 Gif-sur-Yvette Cedex, France
     \and Institut d'Astrophysique Spatiale, CNRS/Universit\'e Paris-Sud 11, 91405 Orsay, France
     \and Canadian Institute for Theoretical Astrophysics - CITA, University of Toronto, 60 St. George Street, Toronto, Ontario, M5S 3H8, Canada
         \and Universit\'e de Bordeaux, Bordeaux, LAB, UMR 5804, 33270, Floirac, France
         \and CNRS, LAB, UMR 5804, 33270, Floirac, France
         \and IAPS- Instituto di Astrofisica e Planetologia Spaziali, via Fosso del Cavaliere 100, 00133 Roma, Italy
          \and European Southern Observatory,  Karl Schwarzschild Str. 2, 85748 Garching bei Munchen, Germany\\
              \email{tracey.hill@cea.fr}
         \thanks{}
             }
   \date{October 2012}

 \abstract{
We present APEX/\partemis\ 450\,\um\ continuum observations of RCW\,36 and the adjacent ridge, a high-mass high-column density filamentary structure at the centre of the Vela~C molecular cloud. These observations, at higher resolution than \herschel's SPIRE camera, reveal clear fragmentation of the central star-forming ridge. Combined with PACS far-infrared and SPIRE sub-millimetre observations from the \herschel\ HOBYS project we build a high resolution column density map of the region mapped with \partemis. We extract the radial density profile of the Vela~C ridge which
with a $\sim$\,0.1\,pc central width is consistent with that measured for low-mass star-forming filaments in the \herschel\ Gould Belt survey. Direct comparison with Serpens South, of the Gould Belt 
Aquila complex, reveals many similarities between the two regions. Despite likely different formation mechanisms and histories,  the Vela~C ridge and Serpens South filament share common characteristics, including their filament central widths.
}


   \maketitle
\section{Introduction}\label{sec:intro}

Independent of mass, the formation of a star is a key astrophysical process: while high-mass stars drive galactic formation and ecology, low-mass stars populate their host galaxy and are themselves likely hosts of planetary systems. Although the formation paradigm for low-mass stars has been relatively well established the formation scenario for high mass stars is less clear, as are the processes involved in their formation \citep{mckee07,zinnecker07}.

The \herschel\ Space Observatory \citep[][]{pilbratt10} is providing 
important observational insights into both low- and high-mass star formation in our Galaxy. 
In particular, two \herschel\ key projects focus on low- and high-mass star formation in relatively nearby star-forming complexes: the Gould Belt survey \citep[HGBS, 130\,--\,500\,pc;][]{andre10} 
and HOBYS
\citep[700\,pc\,--\,3\,kpc;][]{motte10}, respectively. 
Combining these two key projects will help to examine the difference between low- and high-mass star formation.

\herschel\ observations in the far-infrared and submillimetre, the crucial regime for studying the birthplaces of stars, 
have revealed most star-forming complexes in our Galaxy to be comprised of filamentary structures \citep{andre10, molinari10}. Star formation proceeds in the densest (``supercritical'' )  of these filaments, which can be clustered into disorganised networks (nests) or into single dominating ridges \citep{hill11}. \citet{arzoum11} found that in low-mass star-forming regions these interstellar filaments could be characterised  by a standard central width, or thickness, of $\sim$\,0.1\,pc.

The Vela~C molecular cloud was recently observed with \herschel\ as part of the HOBYS project \citep{hill11, giannini12, minier12}. 
Vela~C is known to house low-, intermediate- and high-mass star formation \citep{massi03, netterfield09, hill11} and  is thought to be at an early stage in its evolution ($<$\,10$^6$\,yr).  Running through the centre of Vela~C is a ridge, a high-column density ($\sim$\,100\,mag\footnote{Where N$_{H_2}$ = 0.94 \Av\ $\times$ 10$^{21}$\,\cmsq\ mag$^{-1}$ \citep{bohlin78}}) self-gravitating filament, which houses the majority of the high-mass dense cores in the cloud \citep{hill11}.
 Adjacent to this ridge, and at roughly the centre 
of the Vela~C molecular cloud is the RCW\,36 ionising star cluster. \citet{minier12} showed that the ridge results from the ionisation of an initial sheet of molecular gas. Located at 700\,pc, Vela~C is the closest complex in the HOBYS sample, which allows direct comparison with low-mass star-forming regions, such as those targeted by the HGBS project, at comparable spatial resolution.

Here we present  a study of the Vela~C ridge and  RCW\,36 using the \partemis\ camera\footnote{\partemis\ is a prototype for the larger format \artemis\ camera soon to be installed on APEX \citep{talvard10}.} at 450\,\um\ on APEX. \partemis\ is used here to complement the \herschel\ SPIRE bands in the submillimetre (250, 350, 500\,\um), at 
a factor 2--3 higher resolution (e.g. 11.5\arcsec\ compared with 25\arcsec\ at 350\,\um).

\section{Observations and data reduction}\label{sec:obs} 

\begin{figure}
\includegraphics[angle=270,width=0.48\textwidth]{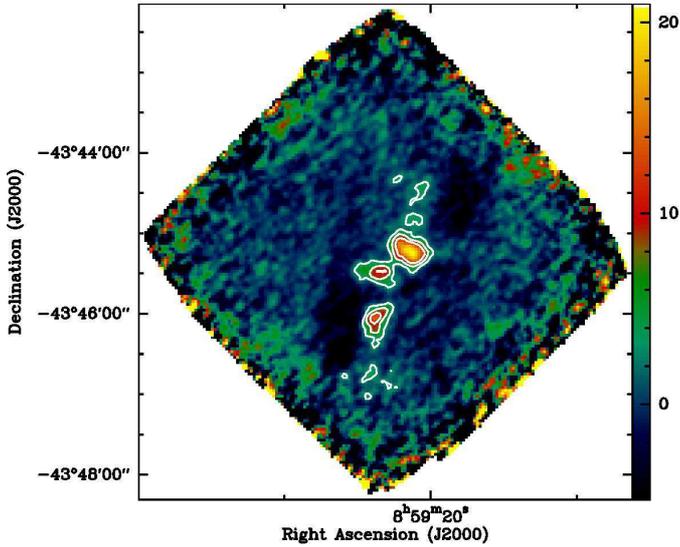}
\caption{\partemis\ 450\,\um\ map of the RCW\,36/Vela~C ridge at 10\arcsec\ resolution. 
The contours are 3, 7, 11 Jy/10\arcsec\ -beam and the rms noise level in the central part 
of the map is $\sim 1\,$Jy/10\arcsec\ -beam.
\label{fig:part}}
\end{figure}

The RCW\,36 region of Vela~C was observed using the \partemis\ bolometer camera \citep{andre08, minier09} on the Atacama Pathfinder Telescope (APEX) telescope on 23 May 2009. 
An area of $\sim$ 4\arcmin\ by 4\arcmin\ was mapped in $\sim 0.5$~h, using a total-power, on-the-fly scanning mode. 
The atmospheric opacity at zenith was measured with a skydip and found to be $\sim$\,0.8 at $\lambda$\,=\,450\,\um, which corresponds to a precipitable water vapour $\sim$\,0.7\,mm. 
The pointing and focus of the telescope were checked using similar observing procedures to those used with APEX/LABOCA (e.g spiral scans). 
Flux calibration was achieved by taking spiral scans of Mars and Saturn.  
The pointing accuracy and absolute calibration uncertainty were estimated to be $\sim$~2\arcsec\  and $\sim$~30\%, respectively.
The main beam had a full width at half maximum (FWHM) $\sim$\,10\arcsec, known to $\sim$\,10\%  accuracy, and 
contained $\sim$\,60\%  of the power, the rest being distributed in an ``error beam'' extending up to an angular radius of $\sim$\,80\arcsec.
The data were reduced using in-house IDL routines, following the same method as \citet{andre08} and \citet{minier09}. 
This procedure includes baseline subtraction, removal of correlated sky noise and 1/\emph{f} noise, and subtraction of uncorrelated 1/\emph{f} noise 
using a method which exploits the high level of redundancy in the data. 
The final 450\,\um\ continuum map is presented in Figure~\ref{fig:part}.

The entire Vela~C molecular cloud ($\sim$\,3\,deg$^2$), including the RCW\,36 region, was mapped with \herschel\ at 70, 160, 250, 350 and 500\,\um\ as part of the HOBYS key program. 
The observations and data reduction are as described by \citet{hill11}.

\section{Structural Analysis}

\begin{figure*}
\begin{center}
\includegraphics[angle=270,width=0.85\textwidth]{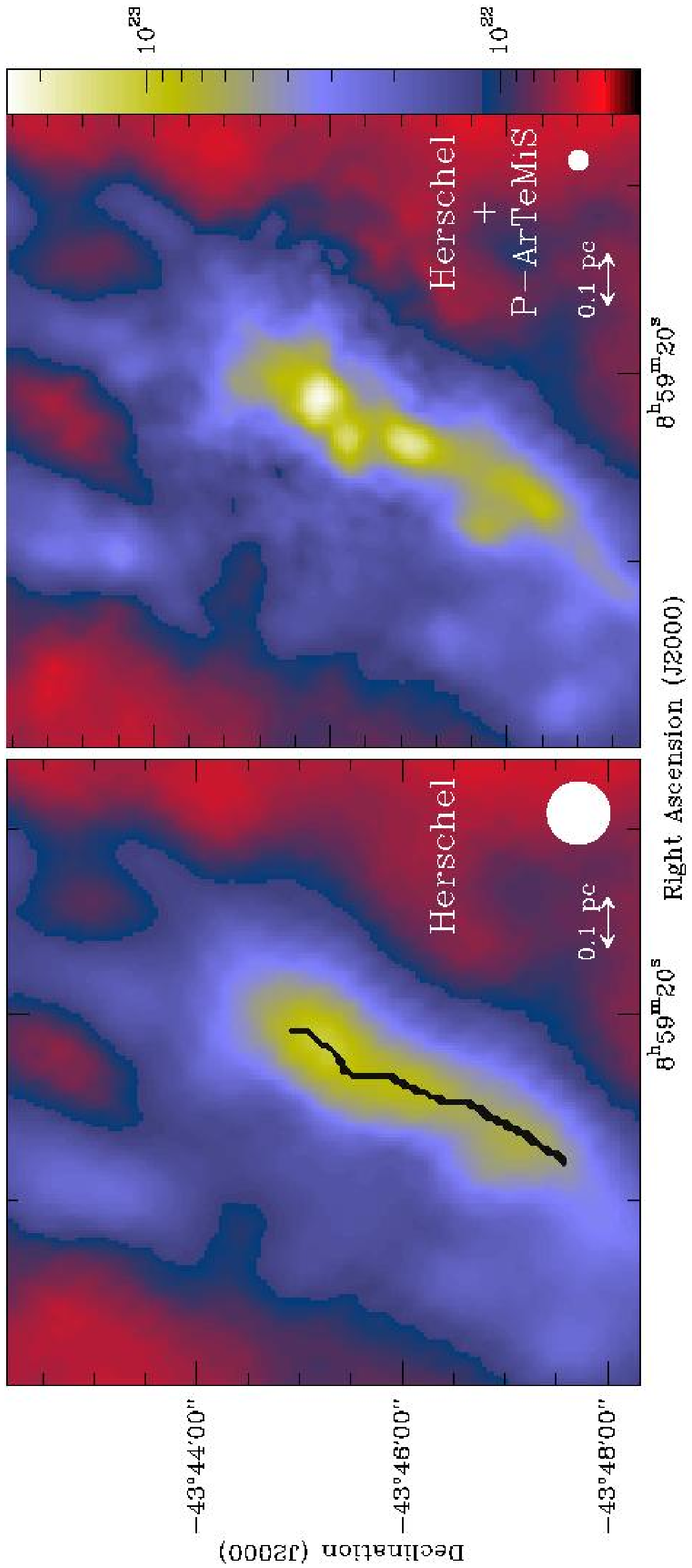}
\caption{The Vela~C ridge/RCW\,36 column density maps. The units are in H$_2\,$ \cmsq. Left: Map
derived from the longest four \herschel\ wavebands with a resolution of 36\arcsec, equivalent to that of the SPIRE 500\,\um\ band.
Right: Map constructed using \partemis\ and \herschel\ data, with resolution 11.5\arcsec. Appendix \ref{sec:cdmap} details how this map was created. 
The filament from which the radial density profile (Fig.~\ref{fig:profile}) was determined (see Section \ref{sec:filaments}) is as indicated by the black line, measured on the right map but, overlaid on the left figure for clarity.
\label{fig:cd}}
\end{center}
\end{figure*}

\subsection{Column density and dust temperature maps}\label{sec:cd}

Multiple wave-band observations covering the far-infrared and sub-millimetre regime allow construction of column density  (N$_{H_2}$)  and dust temperature maps. The maps of these quantities for the Vela~C ridge/RCW\,36 region,  derived from \herschel\ data, were drawn using  pixel-by-pixel spectral energy distribution (SED) fitting according to a modified blackbody with a single dust temperature \citep[cf.][]{hill12}.
Only the longest four \herschel\ wavebands were used.  In order to do this, the \herschel\  observations were first convolved to the resolution of the 500\,\um\ band (36\arcsec), and the zero offsets derived from \Planck\ data were applied \citep[see][]{hill11}. The quality of the SED fit was assessed using $\chi^2$ minimisation. The corresponding  column density map of the Vela~C ridge and RCW\,36 is given in Fig. \ref{fig:cd} (left).

As our \partemis\ data, at 10\arcsec\ resolution have better resolution than that of all of the \herschel\  SPIRE bands, we devised a method to derive a higher resolution column density map of the Vela~C ridge and RCW\,36 
region mapped by \partemis.  By combining the PACS 160\,\um, SPIRE 250\,\um\ and \partemis\ 450\,\um\ data we were able to derive a column density map at 11.5\arcsec\ resolution, i.e. the resolution of the 160\,\um\ \herschel\ map. This method is similar to the one used by \citet{pedro12} for their higher resolution \herschel\ column density map of the Taurus B211 region, but adapted for \partemis\ data as outlined in Appendix \ref{sec:cdmap}.
Due to the area covered by our \partemis\ observations, only
$\sim$\,1/3 of the full Vela~C ridge detected by \herschel\  is measured in the higher resolution column density map,
which shows clear fragmentation into a number of cores/clumps.
These cores are clearly visible with \partemis\ (see Fig.~\ref{fig:part}) but could not be previously identified from the lower resolution \herschel\ column density map (Fig. \ref{fig:cd}, left).

\subsection{Filamentary structure}\label{sec:filaments}

At the 36\arcsec\ resolution of the \herschel\ column density map, the RCW\,36 region is characterised by a single dominating high-column density filament containing the Vela~C ridge (see Fig. \ref{fig:cd}). The filamentary structure seen in our higher resolution column density map is consistent with that seen in the lower resolution map,  with slight deviations as it traces the topological structure of the fragmented cores (see Fig. \ref{fig:cd}, right).
The mean column density of the Vela~C ridge, as measured from the higher resolution column density map, is $\sim$\,9$\times$\,10$^{22}$\,\cmsq.

\begin{figure}
\includegraphics[height=0.33\textwidth]{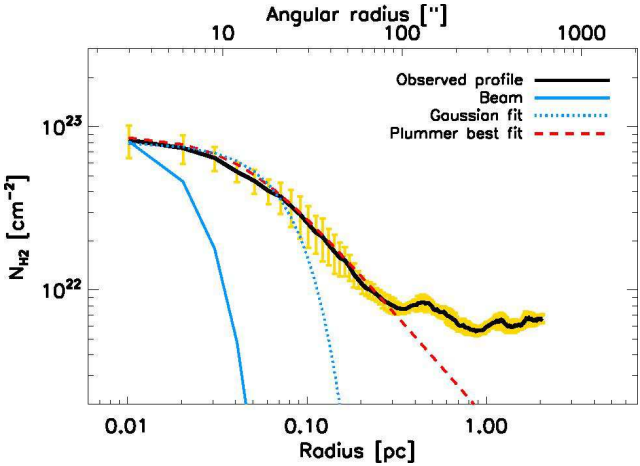}
\caption{The mean radial density profile (western component) perpendicular to the Vela~C ridge (black line on Fig.~\ref{fig:cd}, left), shown here in log-log format. The complementary eastern profile (in the direction of the star cluster) is given in Fig.~\ref{fig:profile:east}.
The area in yellow shows the dispersion of the radial profile along the filament. The solid blue line corresponds to the effective 11.5\arcsec\ HPBW resolution of the column density map (0.04\,pc at 700\,pc) used to construct the profile. The dotted blue line indicates the Gaussian profile, while the dotted red line shows the best fit to the model, often called the Plummer profile. \label{fig:profile}}
\end{figure}

The mean radial density profile ($\rho_p$) of the Vela~C ridge (Fig.~\ref{fig:profile}) was derived by measuring cuts perpendicular to the crest of the filament at each pixel, and then averaging along the length of the filament  \citep[as detailed in][]{arzoum11}. In order to derive the characteristic parameters of the profile (e.g. central density, power law exponent), we assume a cylindrical filament model given by a Plummer-like function, which is a density profile that can be expressed in terms of column density.
Accordingly, 
\begin{equation}
\rho_{p}(r) = \frac{\rho{_c}}{[1 + (r/R_{flat})^2]^{p/2}}
\end{equation}
\noindent
where $\rho_c$ is the radial density at the centre of the filament, $p$ is the exponent of the model function,
 and R$_{flat}$ is the characteristic radius for the flat inner portion of the profile.  

\onlfig{4}{
\begin{figure*}
\includegraphics[height=0.33\textwidth]{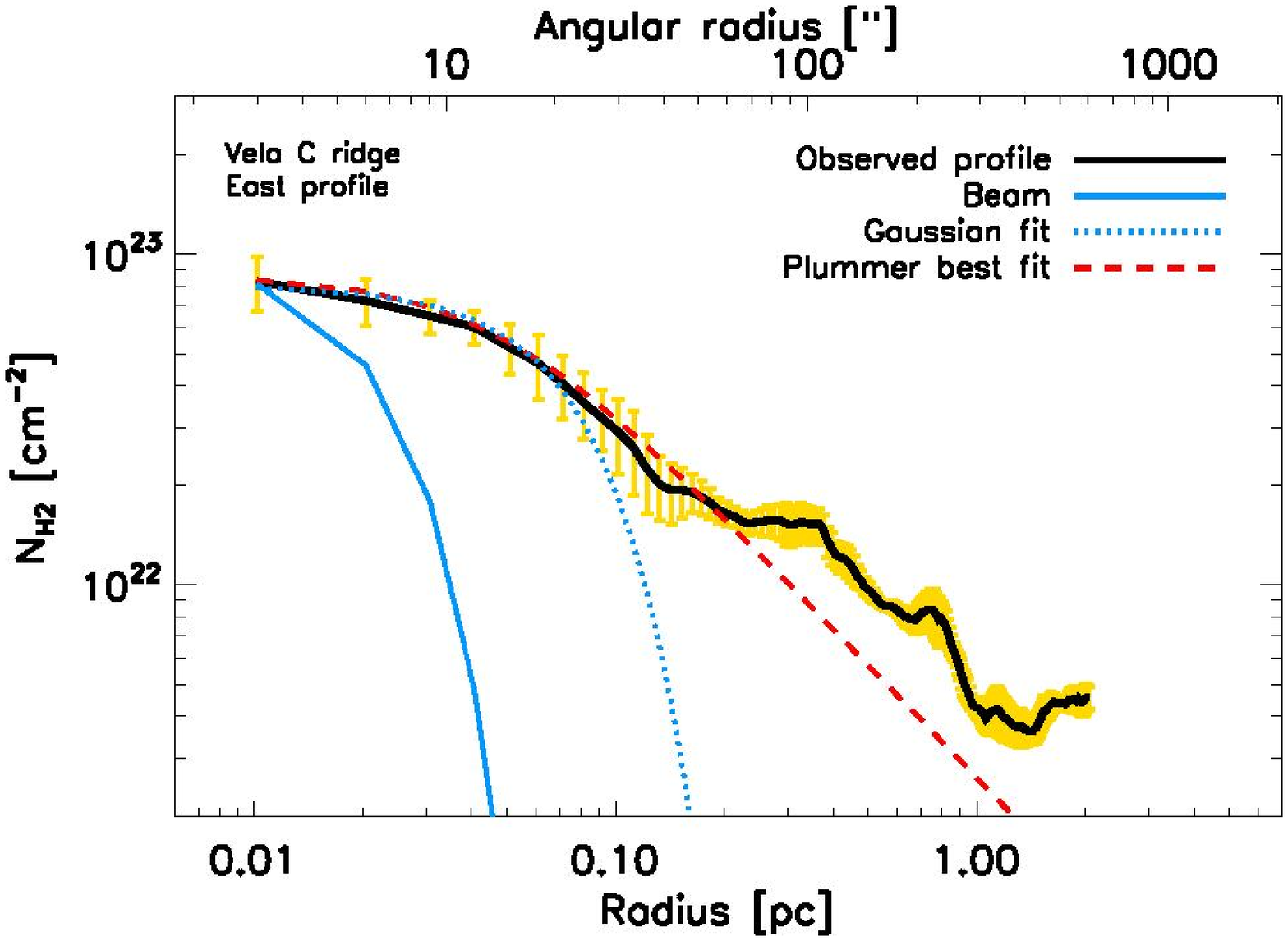}
\includegraphics[height=0.33\textwidth]{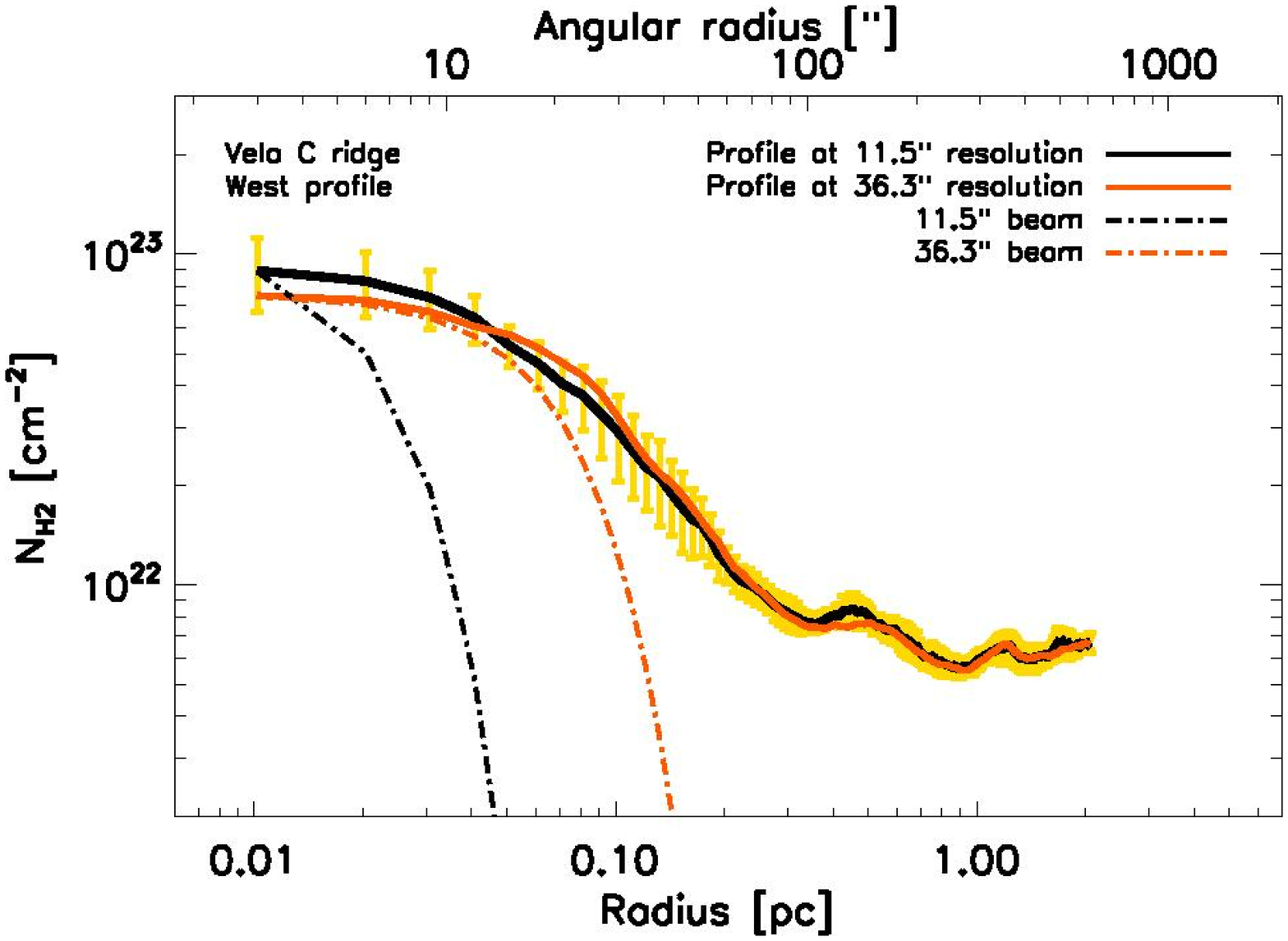}
\caption{Left: Mean radial column density profile measured on the eastern side of the Vela~C ridge in the 11.5\arcsec\ resolution column density map 
shown in Fig.~\ref{fig:cd} (right).
This figure is complementary to Fig.~\ref{fig:profile} for the side containing the OB cluster. This profile has a $p$ value of $2.3 \pm 0.5$ and R$_{flat}$=0.05\,$\pm$\,0.02\,pc.
The error bars in yellow show the dispersion of the radial profile along the filament, while the lines on the plot are as per Fig.~\ref{fig:profile} and as indicated on the key.
Right: Comparison of the mean background-subtracted radial profiles measured on the western side of the Vela~C ridge in the 11.5\arcsec\ resolution column density map (black 
curve and yellow error bars) and in the 36.3\arcsec\ resolution column density map (orange curve). 
The black and orange dash-dotted curves represent the effective 11.5\arcsec\ and 36.3\arcsec\ HPBW resolutions of the corresponding data, respectively.
\label{fig:profile:east}}
\end{figure*}}

The Vela~C ridge can be characterised by an an inner radius R$_{flat} \sim$\,0.05\,$\pm$\,0.02\,pc, and a  deconvolved FWHM of 0.12\,$\pm$\,0.02\,\,pc which is consistent with that seen in low-mass star-forming filaments \citep{arzoum11}. The Vela~C radial density profile decreases at large radii as r$^{-(2.7 \pm 0.2)}$. 
The filament outer radius  $\sim$\,0.4\,$\pm$0.1\,pc is defined from the deviation of the observed (western side) profile from the Plummer fit (cf. Fig. \ref{fig:profile}). The eastern side of the profile, containing the OB star cluster, decreases up to a larger radius $\sim$\,1\,--\,1.5\,pc (see Fig. \ref{fig:profile:east}).

The radial profiles of the column density can also be used to derive a mass per unit length (\mpul\ = $\int \Sigma _{\rm obs}(r)dr$) value for a filament, by integrating the column density over the radius \citep[cf.][]{arzoum11}.
The Vela~C ridge is well defined and constrained allowing us to estimate, without confusion, its mass per unit length
as \mpul ~$\sim$\,320\,($\pm$75) or 400\,($\pm$85) \mstar/pc as measured from the integrated radial profile of 0.4 and 1.5\,pc, respectively, 
after subtracting a background of 3.6\,$\times$\,10$^{21}$\,\cmsq.

\section{Comparison with the Serpens South filament}\label{sec:serpens}

Recent \herschel\ observations have revealed the prolificity of interstellar filaments in star-forming complexes, though only those filaments above an \Av\ of $\sim$\,8\,mag are supercritical and capable of forming stars \citep{andre10,andre11}. \citet{arzoum11} showed that the filaments in low-mass HGBS regions tend to have inner widths of $\sim$\,0.1\,pc.
With higher resolution \partemis\ data at hand it is possible to  check the application of such a characteristic filament width to  
more distant, higher mass regions, such as the Vela~C complex at $d \sim 700$~pc. Here we have shown that, based on its radial column density profile, the Vela~C ridge has a filament inner width consistent with the characteristic inner width suggested by \citet{arzoum11}. 
We have extended our study of filamentary structures to the Serpens South filament, part of the Aquila star-forming complex at $d \sim260$~pc
(see Fig. \ref{fig:serpens}).
The Serpens South filament is particularly interesting as it, as one of the most extreme column density filaments of the HGBS, is comprised of high-column density material, similarly to Vela~C.
\citet{bontemps10} detected seven Class~0 protostars, 
confirming that Serpens South is undergoing low- to intermediate-mass star formation, while \citet{maury11} used evolutionary tracks to estimate the lifetime of these Class-0 protostars 
and showed that Serpens South is at a very early phase of forming stars.

The column density map of Serpens South has been derived in the same manner as that of Vela~C \citep[see section \ref{sec:cd} and][]{konyves10}, with only the longest four \herschel\ bands. The 36\arcsec\ resolution of this map, at the distance of Serpens South, corresponds to a spatial resolution of 0.05\,pc on the sky, comparable to that of Vela~C when using our higher-resolution column density maps with \partemis\ data (0.04\,pc at 700\,pc). These two regions - Serpens South, forming low- to intermediate-mass stars, and Vela~C forming intermediate and potentially high-mass stars - provide 
a good comparative opportunity.

\onlfig{5}{
\begin{figure*}
\includegraphics[height=0.45\textwidth]{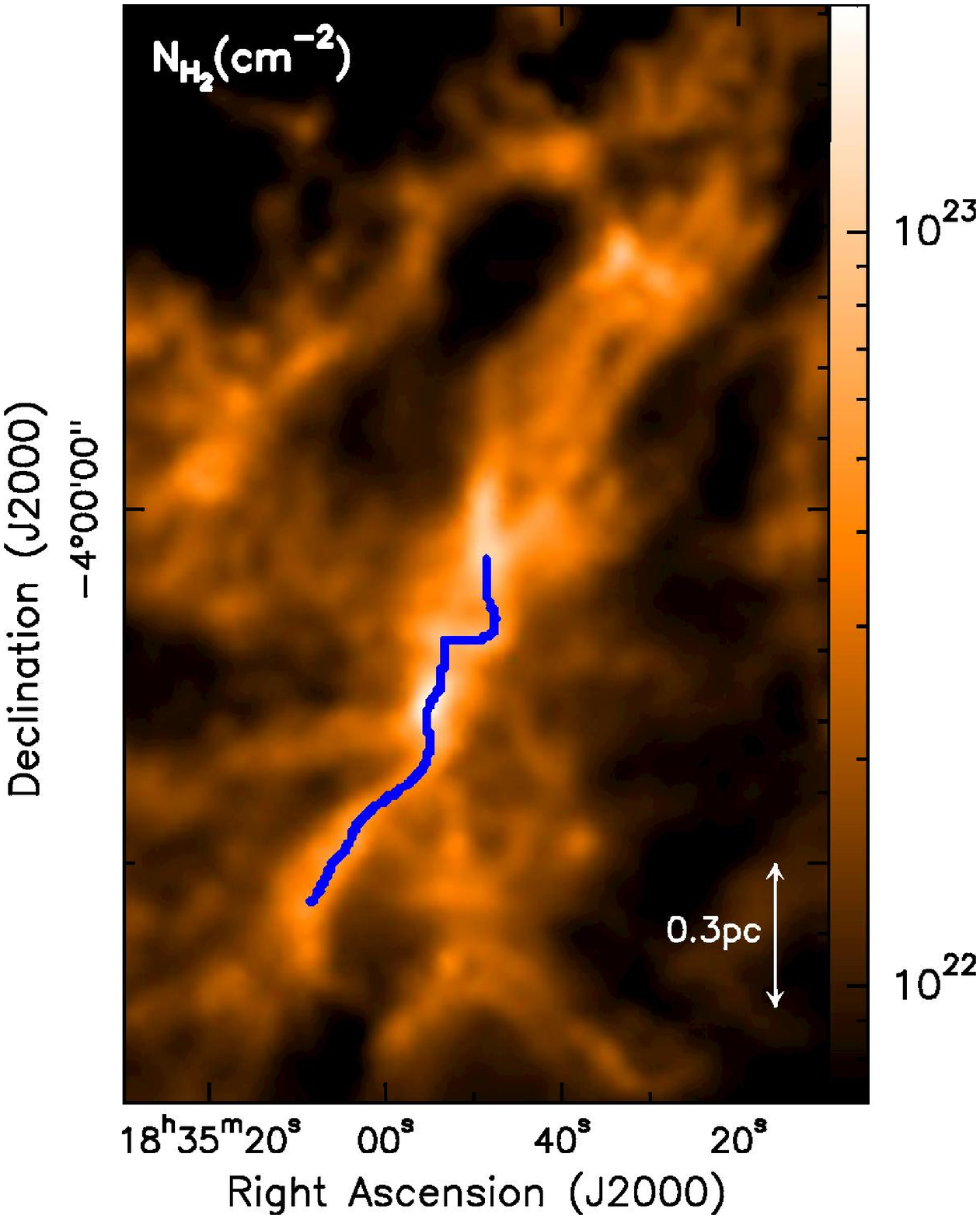}
\includegraphics[height=0.35\textwidth]{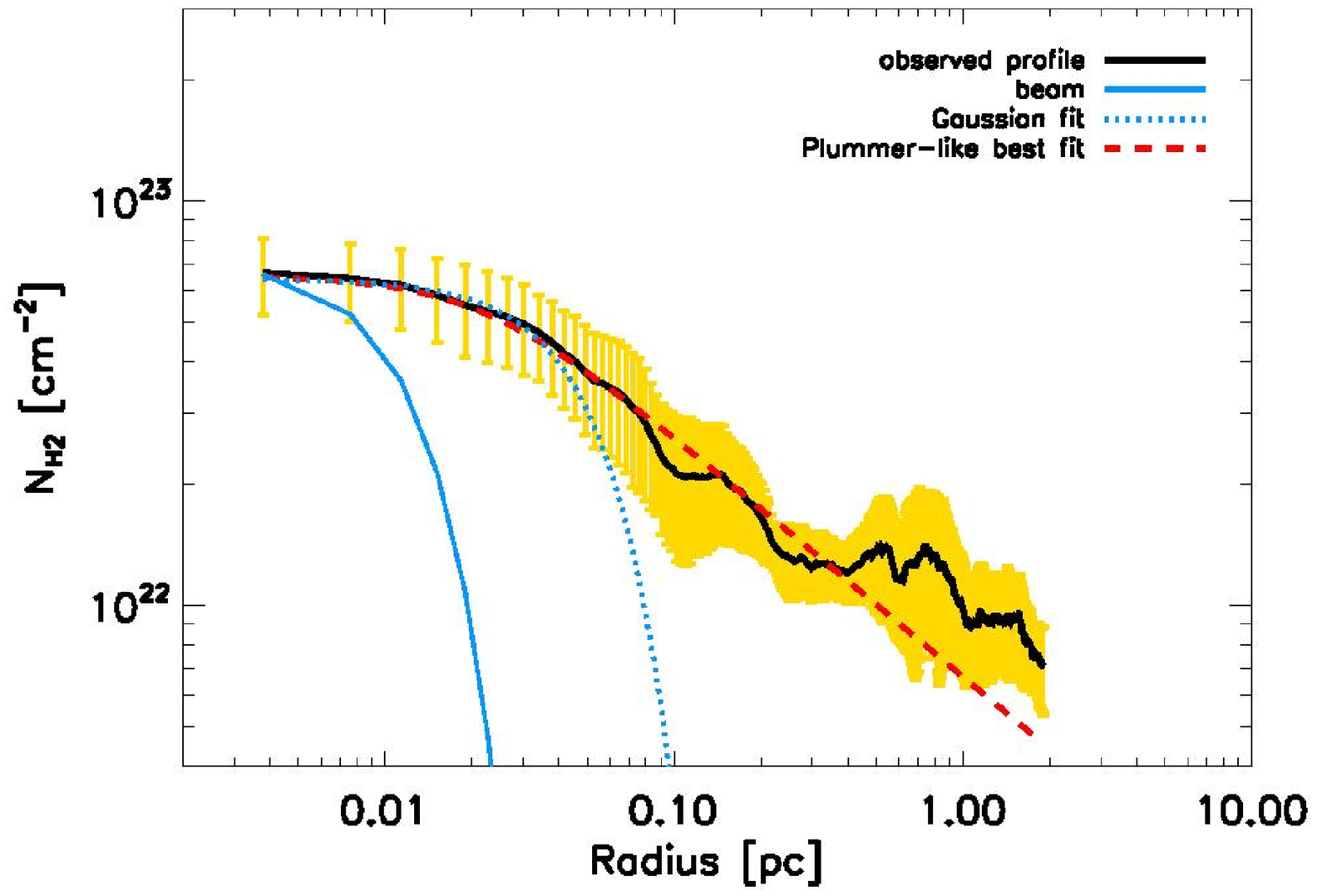}
\caption{Left: Column density map of the Serpens South filament (36\arcsec~resolution)  derived from HGBS data \citep[see][]{konyves10}, with the corresponding topological filament identified overlaid in blue. 
Right: Mean radial density profile (taken from both sides of the filament) measured perpendicular to the supercritical Serpens South filament (left) shown here in log-log format.  Lines and colour coding are consistent with Figs. \ref{fig:profile} and \ref{fig:profile:east}. The Gaussian fit to the inner part of the profile (dotted blue curve) has a deconvolved FWHM width $0.10 \pm 0.05 $~pc.
The best Plummer-like model fit (dashed red curve) has an inner radius R$_{flat} \sim$\,0.03\,$\pm$\,0.01\,pc and a power-law index $p = 2.02 \pm 0.27 $. 
 \label{fig:serpens}}
\end{figure*}}

The crest shown in Fig. \ref{fig:serpens} (left) 
for the Serpens South filament is $\sim$\,1.2\,pc in length and has an average column density value of 6.4\,$\times$\,10$^{22}$\,\cmsq\ (after subtracting a background of 3.7\,$\times$\,10$^{21}$\,\cmsq). The crest shown in Fig. \ref{fig:cd} for 
the Vela~C ridge  is $\sim$\,0.8\,pc in length 
and is slightly more dense at 8.6\,$\times$\,10$^{22}$\,\cmsq. The Vela~C ridge and Serpens South filament have a similar mass per unit length (\mpul\,= 320 and 290 \mstar/pc, respectively) and a similar outer radius ($\sim$\,0.4\,pc;  though the outer radius for both regions may be as large as $\sim$\,1\,--\,1.5\,pc, see Figs. \ref{fig:profile:east} and \ref{fig:serpens}).
The radial density profile of the Serpens South filament  (Fig. \ref{fig:serpens}) has a (deconvolved) 
Gaussian FWHM width of 0.10\,$\pm$\,0.05\,pc, in agreement with that found here for Vela~C and the characteristic width 
of other nearby interstellar filaments measured by \citet{arzoum11} using HGBS data. 
The inner width values measured for the Vela~C ridge and Serpens South filament remain almost unchanged after removing the cores within them:
the deconvolved FWHM values become 0.11\,$\pm$ 0.01\,pc and 0.14 $\pm$ 0.01\,pc, respectively.
As such, the common inner width of $\sim$\,0.1\,pc found in both cases may not be attributed to core properties.

The total length and mass of the Serpens South filament are of the order 2\,pc and 500\,\mstar, respectively, 
while the total length and mass of the Vela~C ridge are $\sim$4\, pc and $\sim $\,600\,\mstar.  
It should also be noted that Vela~C is 
more distant than
Serpens South, and thus core surveys in Vela~C are intrinsically biased toward more massive objects than those accessible in Serpens South.

\section{Universality of star-forming filament profiles?}

Star formation requires a reservoir of material, concentrated into a small volume, from which the burgeoning young star can accumulate mass. The idea of a minimum mass, or density requirement, for star formation inside molecular clouds is then not surprising. 
\citet{evans08} suggested that star formation is restricted to dense gas within molecular clouds, which also has a higher star formation efficiency than lower density gas.
Comparing \spitzer\ inventories of young stellar objects with dust extinction (\Av) maps of nearby molecular clouds, both \citet{lada10} and \citet{heiderman10} suggested that star formation requires a minimum gas density (corresponding to \Av\ $\approx$\,8\,mag). Essentially the same column density threshold was recently found by \citet{andre10, andre11} from an analysis of the prestellar core population in the Aquila complex
based on  \herschel\ data (see also Section \ref{sec:intro}).

In addition to a minimum mass/density for core and star formation, a minimum threshold for high-mass star formation has also been suggested both theoretically \citep{krumholz08} and observationally \citep{kauff10}. According to these authors, low- and high-mass stars likely do not form in the same environments, with the latter requiring a minimum density, and the two modes of star formation (low- and high-mass) are distinct from each other.

In this letter we have compared two nearby ($<$\,700\,pc) regions undergoing clustered star formation. The Vela~C molecular cloud is known as a low- to intermediate-mass star-forming molecular cloud. The Vela~C ridge however is associated with a high-mass  ionising star cluster \citep{minier12}, and hosts seven high-mass dense cores \citep{hill11} each with the potential to form high-mass stars. 
Ridges are the extreme density filaments of high-mass star-forming regions whose large areas of influence suggest that they may 
have been formed through dynamic scenarios such as converging flows, filament mergers and/or ionisation pressure from nearby star clusters \citep{hill11, quang11, hennemann12,minier12}. 
Comparatively, the Serpens South filament is populated by low- to intermediate-mass protostars \citep{gutermuth08, maury11}. 
Both of these filamentary star-forming clumps have supercritical masses per unit length and are thus likely to form more stars in the future.

Our analysis of the filamentary structure in the Vela~C ridge and the Serpens South filament indicates that the two are quite similar with respect to their column density profile 
and mass per unit length, yet the Vela~C ridge is a factor of $\sim$\,2 longer and is slightly more massive 
than the Serpens South filament (see Sect.~\ref{sec:serpens}). 
The lack of known Class-0 protostars in Vela~C  may arise from sensitivity limitations, but the same can not be said for the lack 
of high-mass dense cores in Serpens South. 

While the Vela~C ridge and Serpens South filament are likely to have formed in different environments, 
under different conditions, and to have experienced different histories (with, e.g., the strong ionizing effect of a massive star cluster
in the case of Vela~C, and no such effect in Serpens South), 
these two regions display a similar filament inner width $\sim$\,0.1\,pc which is also consistent with the characteristic inner width 
found for low-mass star-forming filaments by \citet{arzoum11}.
This suggests that 
supercritical filaments and ridges, regardless of their formation process, have the same inner width which may be characteristic across modes of star formation.
These results advocate 
a high degree of commonality among star-forming filaments, independently of the masses of the stars they form, 
rather than the aforementioned idea of distinct environmental conditions for higher mass stars.

More work would be needed to establish the universality of star-forming filament profiles in the high-mass regime. 
Recently, \citet{hennemann12} used HOBYS data to show that the high-mass DR\,21 ridge in the Cygnus X complex ($d \sim 1.4$~kpc) 
has an apparent mean central width of $\sim$\,0.3\,pc when observed at a resolution 0.17\,pc. 
The flat inner portion of the DR\,21 ridge was however only marginally resolved with \herschel\ observations.
It should be stressed that higher mass star-forming ridges and filaments occur at greater distances than lower mass ones, and observations of these are thus subject to lower spatial resolution. At distances exceeding that of Vela~C  ($>$\,700\,pc) a characteristic inner width of $\sim$\,0.1\,pc would remain unresolved, or at best marginally resolved with \herschel.
Higher resolution observations of high-mass star-forming filaments and ridges, with for example the full 
\artemis\ camera to be installed soon on APEX or the Atacama Large Millimetre Array (ALMA), over a greater number of regions 
are needed now to address the hypothesis of a characteristic inner width for interstellar filaments independently of the masses of the stars they form.

\begin{acknowledgements}
T.H. is supported by a CEA/Marie-Curie Eurotalents Fellowship.
 Part of this work was supported by the ANR (\emph{Agence Nationale pour la Recherche}) project `PROBeS', number ANR-08-BLAN-0241.

\end{acknowledgements}

\bibliographystyle{aa} \bibliography{/Users/thill/pap_write/bib/references} 
\begin{appendix}

\section{Deriving a high-resolution column density map with \partemis\ and \herschel\ data}\label{sec:cdmap}

The \partemis\ 450\,$\mu$m map (Fig.~\ref{fig:part}) provides information on small-scales in the RCW~36 ridge but is not 
sensitive to large-scale structures because any emission at low spatial frequencies has been completely filtered out during the process of atmospheric skynoise removal. 
Conversely, the \herschel\ column density map produced by \citet{hill11} at 36\arcsec\ resolution contains little information on scales $<$ 36\arcsec\ 
but provides an accurate view of larger-scale features.
In order to combine these two complementary data sets, we used the following procedure \citep[inspired from][]{pedro12}.

Following the spirit of a multi-resolution data decomposition \citep[cf.][]{starck06}, 
the gas surface density 
distribution of the RCW\,36 region, smoothed to the resolution of the \herschel/PACS 160\,$\mu$m observations, may be expressed as a sum of 
four terms:  $$ \Sigma_{160} =  \Sigma_{500} +  \left(\Sigma_{350} - \Sigma_{500}\right) +  \left(\Sigma_{250} - \Sigma_{350}\right) +  \left(\Sigma_{160} - \Sigma_{250}\right). \ \ (A.1)$$ 

\noindent
where $\Sigma_{500}$,  $\Sigma_{350}$, $\Sigma_{250}$, and $\Sigma_{160}$ represent smoothed versions of the 
intrinsic gas surface density distribution $\Sigma$ after convolution with the \herschel\ beam at 500\,$\mu$m, 350\,$\mu$m, 250\,$\mu$m, and 160\,$\mu$m
respectively, i.e.:  $\Sigma_{500} = \Sigma * B_{500}$, $\Sigma_{350} = \Sigma * B_{350} $, $\Sigma_{250} = \Sigma * B_{250}$, and $\Sigma_{160} = \Sigma * B_{160}$.

The first term of Eq.~(A.1)  is simply the surface density distribution smoothed to the resolution of the \herschel/SPIRE 500\,$\mu$m data. 
An estimate,  $\bar{\Sigma}_{500} $, of this term can be obtained in a manner similar to \citet{hill11} through pixel-by-pixel SED fitting to the longest four \herschel\ data points, 
assuming the following dust opacity law, very similar to that advocated by 
\citet{hildebrand83} 
at submillimetre wavelengths:  
$\kappa_{\nu} = 0.1 \times (\nu/1000~{\rm GHz})^{\beta} =  0.1 \times (300~{\rm \mu m}/\lambda)^{\beta}$~cm$^2$/g, with $\beta = 2$.

The second term of Eq.~(A.1) may be written as $\Sigma_{350} - \Sigma_{350}*G_{500\_350} $, 
where $G_{500\_350} $ is a circular Gaussian with full width at half maximum (FWHM) $\sqrt{36.3^2- 24.9^2} \approx 26.4\arcsec $. 
(To first  order, the SPIRE beam at 500\,$\mu$m is a smoothed version of the SPIRE beam at 350\,$\mu$m, i.e., 
$B_{500} = B_{350}*G_{500\_350} $.)
The second term of Eq.~(A.1) may thus be viewed as a term adding information on spatial scales accessible 
to SPIRE observations at  $350\,\mu$m, but not at  $500\,\mu$m. 
In practice, one can derive and estimate $\bar{\Sigma}_{350} $ of $\Sigma_{350} $ in a manner similar to $\bar{\Sigma}_{500} $, 
through pixel-by-pixel SED fitting to three $Herschel$ data points between 160\,$\mu$m and 350\,$\mu$m (i.e., ignoring the lower resolution 500\,$\mu$m data point).
An estimate of the second term of Eq.~(A.1) can then be obtained by subtracting a smoothed version of $\bar{\Sigma}_{350} $
(i.e., $\bar{\Sigma}_{350}*G_{500\_350} $) to $\bar{\Sigma}_{350} $ itself, i.e., by removing low spatial frequency information from $\bar{\Sigma}_{350} $.

Likewise, the third term of Eq.~(A.1) may be written as $\Sigma_{250} - \Sigma_{250}*G_{350\_250} $, where 
$G_{350\_250} $ is a circular Gaussian with FWHM $\sqrt{24.9^2-18.2^2} \approx 17.0\arcsec $, and may be 
understood as a term adding information on spatial scales only accessible to $Herschel$ observations at wavelengths $\leq 250 \mu$m. 
In order to derive an estimate $\bar{\Sigma}_{250} $ of $\Sigma_{250} $ on the right-hand side of Eq.~(A.1), 
we first smoothed the PACS 160\,$\mu$m map to the 18.2\arcsec\ resolution of the SPIRE 250\,$\mu$m 
map and then derive a color temperature map between 160\,$\mu$m and 250\,$\mu$m from the observed $I_{\rm 250 \mu m }(x,y)/I_{\rm 160 \mu m }(x,y)$ 
intensity ratio at each pixel $(x,y)$. The SPIRE 250\,$\mu$m map was converted to a gas surface density map ($\bar{\Sigma}_{250} $), assuming 
optically thin dust emission at the temperature given by the color temperature map and a dust opacity at 250\,$\mu$m 
$\kappa_{\rm 250 \mu m } = 0.1 \times (300/250)^2$~cm$^2$/g. 
An estimate of the third term of Eq.~(A.1) can then be obtained by subtracting a smoothed version of $\bar{\Sigma}_{250} $ 
(i.e., $\bar{\Sigma}_{250} *G_{350\_250} $) to $\bar{\Sigma}_{250} $ itself, i.e., by removing low spatial frequency information from $\bar{\Sigma}_{250} $. 

The fourth term on the right-hand side of Eq.~(A.1) is the component containing information on the smallest scales $\sim$\,10\,--\,18\arcsec\ 
and which was estimated using the \partemis\  data.
It may be written as $\Sigma_{160} - \Sigma_{160}*G_{250\_160} $, where 
$G_{250\_160} $ is a circular Gaussian with FWHM $\sqrt{18.2^2-11.5^2} \approx 14.1\arcsec $.  
In order to derive an estimate $\bar{\Sigma}_{160} $ of $\Sigma_{160} $ on the right-hand side of Eq.~(A.1), 
we used the color temperature map between 160\,$\mu$m and 250\,$\mu$m (18\arcsec\ resolution) to 
convert a slightly smoothed version of the \partemis\ 450\,\um\ data (11.5\arcsec\ resolution) and the PACS 160\,$\mu$m data 
to a column density map at 11.5\arcsec\ resolution, assuming optically thin dust emission and the same dust opacity law as above.
Due to skynoise filtering, the latter map ($\bar{\Sigma}_{160} $)  does not contain information on angular scales $\ga 40\arcsec $, but this is not a problem 
since our estimate of the fourth term of Eq.~(A.1) was obtained by subtracting a smoothed version of $\bar{\Sigma}_{160} $ 
(i.e., $\bar{\Sigma}_{160} *G_{250\_160} $) to $\bar{\Sigma}_{160} $ itself, i.e., by removing any information on scales $\ga 18\arcsec $ from $\bar{\Sigma}_{160} $. 
This subtraction of scales larger than $\sim18\arcsec $ from $\bar{\Sigma}_{160} $ 
also suppresses the effect of the bowls of negative emission seen at $\pm 30\arcsec$ on either side of the central ridge in 
the  \partemis\ 450\,\um\ map (Fig.~\ref{fig:part}).

Our final estimate $ \tilde{\Sigma}_{160}$ of the gas surface density distribution at 11.5\arcsec ~resolution was produced by summing 
the above estimates of the four terms on the right-hand side of Eq.~(A.1): 
$$ \tilde{\Sigma}_{160} =  \bar{\Sigma}_{500} +  \left(\bar{\Sigma}_{350} - \bar{\Sigma}_{350}*G_{500\_350}\right) 
+  \left(\bar{\Sigma}_{250}  - \bar{\Sigma}_{250}*G_{350\_250} \right) $$
$$ +  \left(\bar{\Sigma}_{160}  - \bar{\Sigma}_{160}*G_{250\_160} \right). \ \  (A.2)$$

\begin{figure}
\includegraphics[angle=270,width=0.4\textwidth]{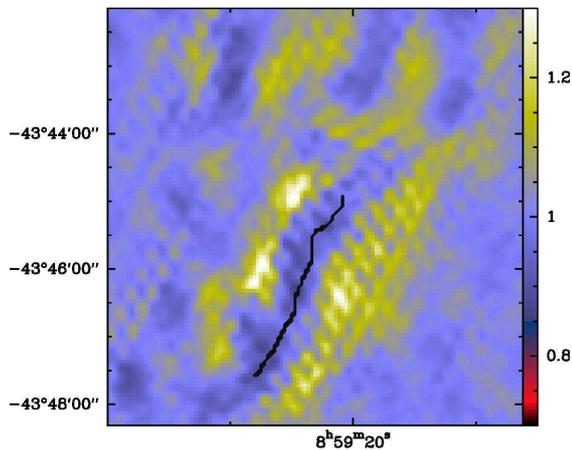}
\caption{Ratio map of the 11.5\arcsec\ resolution column density map (corresponding to $ \tilde{\Sigma}_{160}$) smoothed to 36.3\arcsec\ resolution divided by 
the standard column density map (corresponding to $\bar{\Sigma}_{500} $). The same filament crest as in Fig.~\ref{fig:cd} is overlaid. 
The maximum value of the ratio is 1.35 and the minimum value is 0.85. 
\label{fig:ratiomap}}
\end{figure}

\noindent
The resulting 11.5\arcsec\ resolution column density map $\tilde{N}_{\rm H_{2}}$
for the RCW36 region is displayed in the right panel of Fig.~2 
in units of mean molecules per $\rm{cm}^{2}$, where $ \tilde{\Sigma}_{160}  = \mu \, {\rm m_H} \, \tilde{N}_{\rm H_{2}}$ and $\mu = 2.33$ is the mean molecular weight.
This high-resolution column density map is subject to larger uncertainties than the standard 36.3\arcsec -resolution column density map 
derived from \herschel\ data. In particular, the absolute calibration uncertainty of the ground-based \partemis\ 450\,\um\ data is larger 
($\sim$\,30\%) than that of the \herschel\ data ($\la$\,10\%), and the beam shape of the \partemis\ instrument is also more 
uncertain\footnote{Note, however, that the $\la$ 80\arcsec\ error beam of the \partemis\ instrument is effectively filtered out during the data reduction 
and map reconstruction process, so that only the $\sim$\,10\% uncertainty in the $\sim$\,10\arcsec\ main beam of \partemis\ matters here for $\tilde{\Sigma}_{160}$.}
than the beams of the SPIRE and PACS cameras on \herschel. 
To test the reliability and robustness of this 11.5\arcsec\ resolution column density map, we smoothed it to the 36.3\arcsec\ resolution of 
the standard column density map (corresponding to $\bar{\Sigma}_{500}$) and inspected the ratio map between the two, which has a mean value of 1.00 and 
a standard deviation of 0.03 (see Fig.~\ref{fig:ratiomap}). 
The two column density maps agree to better than $35\%$ everywhere. 
The morphology and amplitude of the features seen in the ratio map (Fig.~\ref{fig:ratiomap}) suggest that the potential artefacts 
present in the high-resolution column density map translate into an uncertainty of $\la$ 40\%  
in the radial column density profiles shown in Fig.~\ref{fig:profile} and Fig.~\ref{fig:profile:east} (left). 
This uncertainty is comparable to the yellow error bars shown on the profiles (corresponding to the dispersion of the radial profiles observed
along the ridge) and therefore does not significantly affect our conclusions regarding the shape and central width of the radial density profile.

Finally, we note that our column density map of the Vela~C ridge 
is also more uncertain than that of the Serpens South filament 
owing to the strong heating effect of the RCW36 cluster and its uncertain location 
along the line of sight. Simple tests suggest that this effect leads to an additional 
50\% uncertainty in the absolute calibration of the column density map of the Vela C ridge 
but has little influence on the shape of the radial column density profile.

\end{appendix}
\end{document}